# Study of wavelength dependence of mode instability based on a semi-analytical model

Rumao Tao, Pengfei Ma, Xiaolin Wang*, Pu Zhou**, Zejin Liu

***Abstract*—We present theoretical study of wavelength dependence of mode instability (MI) in high power fiber lasers, which employs an improved semi-analytical theoretical model. The influence of pump / seed wavelength and photodarkening on threshold has been studied. The results indicate promising MI suppression through pumping or seeding at an appropriate wavelength. Small amounts of photodarkening can lead to significant impact on MI.**

## I. Introduction

In recent years, fiber laser systems based on large mode area double cladding fibers have rapidly evolved into light sources able to deliver single-mode output powers beyond ten-kilowatt level [1]. However, large mode area, which means multi-mode operation under the current technological limitations, inevitably results in the onset of mode instabilities (MI), which currently limits the further power scaling of ytterbium doped fiber laser systems with diffraction-limited beam quality [2]. Due to the far-reaching impact of MI, the first detailed experimental report of this phenomenon triggered the publication of several theoretical models dealing with it numerically [3-11] or analytically [12-15]. Although fully numerical model can take a large variety of physical effect into consideration and is very useful for precise quantitative analysis [3-11], many aspects of the underlying physics are often lost in the numerical process. Semi-analytical model can be realized at the cost of some accuracy by adopting some approximation [12-15], which can provides a good understanding of physical insight.

Recently, some new experimental phenomena, such as wavelength dependence of the threshold [16-18] and impact of photodarkening [19], have been reported, which has not been explained theoretically until now. Based on an improved semi-analytical steady-periodic model of MI in ytterbium doped fiber laser, the influence of various pump /seed wavelengths, photodarkening on MI has been studied, and the aforementioned experimental phenomena can be understood at the light of the theoretical results presented in this paper.

Manuscript received Febuaray 17, 2015. This work was supported in part by the National Science Foundation of China under grant No. 61322505, the program for New Century Excellent Talents in University, the Innovation Foundation for Excellent Graduates in National University of Defense Technology under grant B120704 and Hunan Provincial Innovation Foundation for Postgraduate under grant CX2012B035.

Rumao Tao, Pengfei Ma, Xiaolin Wang, Pu Zhou and Zejin Liu are with College of Optoelectric Science and Engineering, National University of Defense Technology, Changsha, Hunan 410073, China (e-mail: chinawxllin@ 163.com, zhoupu203@ 163.com).

## II. Theoretical model

The physical principle of MI is thought to be stimulated thermal Rayleigh scattering [5, 12], which is first elaborated by Smith *et al.* in [3]. This physical principal has been employed widely in theoretical study of MI [4-15], and agrees with the experimental observation. So this physical principle is employed as the physical basis of our theoretical model in [20]. However, the heat due to the absorption in the fiber has not been taken into consideration in [20]. Assuming that all the power absorbed due to linear absorption is turned into heat, the volume heat-generation density $Q$ can be approximately expressed as

$$Q(r,\phi,z,t)\cong g(r,\phi,z,t)\left(\frac{\nu_p-\nu_s}{\nu_s}\right)I_s(r,\phi,z,t)+\gamma(r,\phi)I_s(r,\phi,z,t) \tag{1}$$

and the gain of the amplifier $g(r,\phi,z,t)$ and the steady state population inversion fraction $n_u$ [3] are given by

$$g(r,\phi,z,t)=\left[\left(\sigma_s^a+\sigma_s^e\right)n_u(r,\phi,z,t)-\sigma_s^a\right]N_{Yb}(r,\phi) \tag{2a}$$

$$n_u(r,\phi,z,t)=\frac{P_p(z,t)\sigma_p^a/h\nu_pA_p+I_s(r,\phi,z,t)\sigma_s^a/h\nu_s}{P_p(z,t)\left(\sigma_p^a+\sigma_p^e\right)/h\nu_pA_p+I_s(r,\phi,z,t)\left(\sigma_s^a+\sigma_s^e\right)/h\nu_s+1/\tau} \tag{2b}$$

where $\nu_{p(s)}$ is the optical frequencies, $\sigma_s^a$ and $\sigma_s^e$ are the signal absorption and emission cross sections, $\sigma_p^a$ and $\sigma_p^e$ are the pump absorption and emission cross sections, $N_{Yb}(r,\phi)$ is the doping profile, $P_p$ is the pump power, $A_p$ is the area of the pump cladding, and $\tau$ is the ion upper-state lifetime. The linear absorption coefficient $\gamma(r,\phi)$ can be non-uniform in ($x$, $y$) to accommodate a photo-darkening model. Similar to the derivation in [20], we can obtain the nonlinear coupling coefficient

$$\chi(\Omega)=2\frac{n_0\omega_2^2}{c^2\beta_2}\mathrm{Im}\left[\iint\left(\bar{h}_{12}+\bar{h}_{12}'\right)\psi_1\psi_2 r\,dr\,d\phi\right] \tag{3}$$

with

$$\bar{h}_{kl}(r,\phi,z)=\frac{\alpha n_2}{\pi}\sum_{\nu}\sum_{m=1}^{\infty}\frac{R_\nu(\delta_m,r)}{N(\delta_m)}\frac{B_{kl}(\phi,z)}{\alpha\delta_m^2-j\Omega} \tag{4a}$$

$$\bar{h}_{kl}'(r,\phi,z)=\frac{\eta\alpha}{\pi\kappa}\sum_{v}\sum_{m=1}^{\infty}\frac{R_v(\delta_m,r)}{N(\delta_m)}\frac{B_{kl}'(\phi,z)}{\alpha\delta_m^2-j\Omega} \tag{4b}$$

$$B_{kl}(\phi,z)=\int_0^{2\pi}d\phi'\int_0^R g_0R_v(\delta_m,r')\cos v(\phi-\phi')\frac{\psi_k(r',\phi')\psi_l(r',\phi')}{(1+I_0/I_{saturation})^2}dr' \tag{4c}$$

$$B_{kl}'(\phi,z)=\int_0^{2\pi}d\phi'\int_0^R\gamma(r',\phi')R_v(\delta_m,r')\cos v(\phi-\phi')\psi_k(r',\phi')\psi_l(r',\phi')dr' \tag{4d}$$

$$\frac{1}{N(\delta_m)}=\frac{1}{\int_0^R rR_v^2(\delta_m,r)dr} \tag{4e}$$

$$n_2=\frac{\eta}{\kappa}\left(\frac{v_p-v_s}{v_s}\right) \tag{4f}$$

$$\Omega=\omega_1-\omega_2 \tag{4g}$$

The meaning of the symbols is the same as in [20], where $\alpha=\kappa/\rho C$, $\rho$ is the density, $C$ is the specific heat capacity, $\kappa$ is the thermal conductivity, $\eta$ is the thermal-optic coefficient, and $\omega$ is the angular frequency of the mode. $R_v(\delta_m,r)$ is given by $R_v(\delta_m,r)=J_v(\delta_m r)$ and $\delta_m$ is the positive roots of $\delta_m J_v'(\delta_m R)+h_q J_v(\delta_m R)/\kappa=0$ ($h_q$ is the convection coefficient for the cooling fluid). Here steady-periodic assumption is used to achieve the above expression, which employed widely in [4-7, 11-15] and corresponds well with reported behavior of amplifiers operating near the thresholds [7].

As can be seen from Eq. (3), the nonlinear coupling between fundamental mode and high order mode is related to thermal-optical effect ($\eta$), quantum defect, gain saturation and the overlap integral of optical fields, which implicate that MI can be suppressed through changing the fiber host materials with a higher thermal conductivity and/or smaller thermo-optic coefficients [21], reducing quantum defect by detuning pump and/or seed wavelength or by tandem pumping [2, 22], reducing the core-to-cladding ratio [16, 23], and gain tailoring [24]. We can also obtain that, if the frequency offset $\Omega$ is zero, the nonlinear coupling coefficient will be zero and no mode coupling is possible [3]. According to [20], the high order mode fraction for the quantum noise induced MI is given as

$$\xi(L)\approx\frac{\hbar\omega_0}{P_1(L)}\sqrt{\frac{2\pi}{\int_0^L P_1(z)\left|\chi''(\Omega_0)\right|dz}}\times\exp\left\{\int_0^L\left[\iint g(r,\phi,z)\psi_2\psi_2 rdrd\phi\right]dz+\int_0^L P_1(z)\chi(\Omega_0)dz\right\} \tag{5a}$$

where $L$ is the length of the gain fiber, $\Omega_0$ denotes the frequency of the maximum of $\chi$. $\chi''$ denotes the second derivative of $\chi$ with respect to $\Omega$. For the case that MI is seeded by intensity noise in the input signal, the high order mode fraction can be expressed as

$$\begin{aligned}\xi(L)&\approx\xi_0\exp\left[\int_0^L dz\iint g(r,\phi,z)(\psi_2\psi_2-\psi_1\psi_1)rdrd\phi\right]\\&+\frac{\xi_0}{4}\sqrt{\frac{2\pi}{\int_0^L P_1(z)\left|\chi''(\Omega_0)\right|dz}}R_N(\Omega_0)\exp\left[\int_0^L P_1(z)\chi(\Omega_0)dz\right]\\&\times\exp\left[\int_0^L dz\iint g(r,\phi,z)(\psi_2\psi_2-\psi_1\psi_1)rdrd\phi\right]\end{aligned} \tag{5b}$$

where $\xi_0$ is the initial high order mode content. Replacing the nonlinear coupling coefficient in Eq. (5) with the new form of Eq. 4(a), heat generated from the quantum defect and absorption can be taken into consideration, and the effects related to absorption, such as photodarkening, can be studied.

## III. NUMERICAL RESULTS

In this section, we first compared our calculation with that obtained by the numerical method [25]. The fiber parameters used for the comparison are the same as those in [25], which are listed in Table I. These parameters are typical of high power ytterbium doped amplifiers and all fiber amplifiers are co-pumped. It is pointed out in [25] that changing the core diameter while keeping the core-to-cladding ratio and *NA* fixed, the threshold powers do not change. This conclusion is given without detail results. To verify our model, we calculated the threshold of quantum noise induced MI and maximal coupling frequency as a function of core size while the core-to-cladding ratio and *NA* are fixed, which is shown in Fig. 1. In the calculation, a quantum noise spectral power density of $\hbar\omega$ was employed [14]. As claimed in [25], it indicates that the threshold powers do not change and the coupling frequency changes proportional to $1/A_{core}$. In addition, the threshold power here is 622W corresponding to pump power of 648W, which is different from that given in [25] by an error of about 5%.

TABLE I
Parameters of Test Amplifier

| | |
|---|---|
| $n_{clad}$ | 1.45 |
| *NA* | 0.054 |
| $\lambda_p$ | 976nm |
| $\lambda_s$ | 1032nm |
| $h_q$ | 5000 W/(m$^2$K) |
| $\eta$ | $1.2\times10^{-5}$ K$^{-1}$ |
| $\kappa$ | 1.38 W/(Km) |
| $\rho C$ | $1.54\times10^6$J/(Km$^3$) |
| $N_{Yb}$ | $3.0\times10^{25}$ m$^{-3}$ |
| $\sigma_p^a$ | $2.47\times10^{-24}$m$^2$ |
| $\sigma_p^e$ | $2.44\times10^{-24}$ m$^2$ |
| $\sigma_s^a$ | $5.8\times10^{-27}$ m$^2$ |
| $\sigma_s^e$ | $5.0\times10^{-25}$ m$^2$ |
| $\tau$ | 901 µs |
| $P_1(0)$ | 10W |
| *absorption* | 0 |
| $R_{coil}$ | $\infty$ |

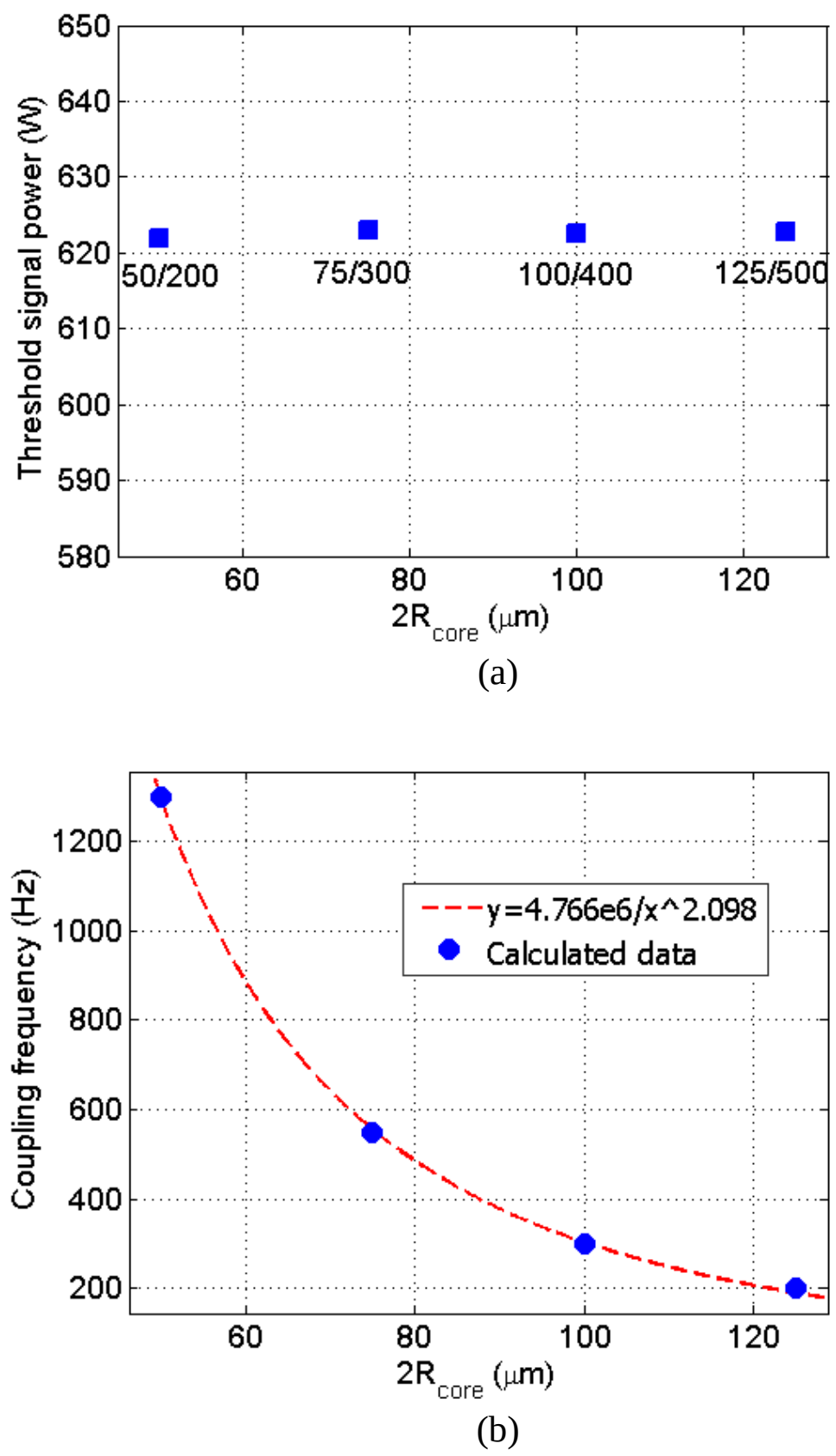

Fig. 1. MI characteristics under various core sizes with the core-to-cladding ratio and *NA* fixed. (a) Threshold power for different core size and (b) different coupling frequency at different core size.

We calculated the threshold and corresponding peak core heat under different pump wavelength in Fig. 2. The lengths of the fibers are adjusted the minimum value necessary to achieve high efficiency, defined as pump absorption ≈0.95. It shows that threshold power is dependent on pump wavelength, and by shifting the pump wavelength from 976nm to 970nm or 985nm, the threshold power can be increased by 40% or 80%, respectively, which agrees with experimental reports [17]. According to Eq. (2b), smaller pump absorption coefficient will lead to lower upper state populations and increasing of spatial hole burning. With increasing of spatial hole burning, the gain saturation is stronger, which results in the change in the transverse heat profile [25]. Then the temperature profile changes, which ultimately reduces the mode coupling by reducing the nonlinear coupling coefficient $\chi(\Omega)$ and increases the MI threshold. So if the pump light is detuned from the absorption peak at 976 nm, the reduced pump absorption coefficient will lead to lower upper state populations, which tends to increase hole burning and thus increase threshold power. In [17], the measured change in threshold by shifting the pump wavelength from 977nm to 970nm was ~78%, which is larger than the calculated 40%. This may be due to that the fiber calculated in our paper has a larger core-clad ratio than that measured in [17], which results in weaken of gain saturation [25] and weaken the suppression capability of MI by shifting the pump wavelength [26]. It can be seen from Fig. 2 (b): although the peak core heat load at 915nm is larger than 920nm, the threshold at 915 is higher than that at 920nm; the peak core heat load at 965nm is smaller than that at 960nm, but the threshold at 965 is lower than that at 960nm; at other pump wavelength, larger peak core heat load indicates lower threshold power. This means that heat load may not have any relation with the MI threshold power as claimed in [17]. Detailed theoretical investigation indicates that the portion of the heat profile that is responsible for mode coupling gain is the antisymmetric part created by the antisymmetric part of the signal irradiance [25], where the antisymmetric part is created due to the high order mode. Under or near the MI threshold, power content of high order mode is far smaller than that of the fundamental mode, which means that the thermal load is mainly composed of symmetric part, which is not responsible for the mode coupling. In addition, it is known that heat load is related to the dopant concentration while MI is independent of that due to that the gain saturation and quantum defect is unchanged as dopant concentration changing [28]. So it is the transverse heat profile instead of the heat load that is important.

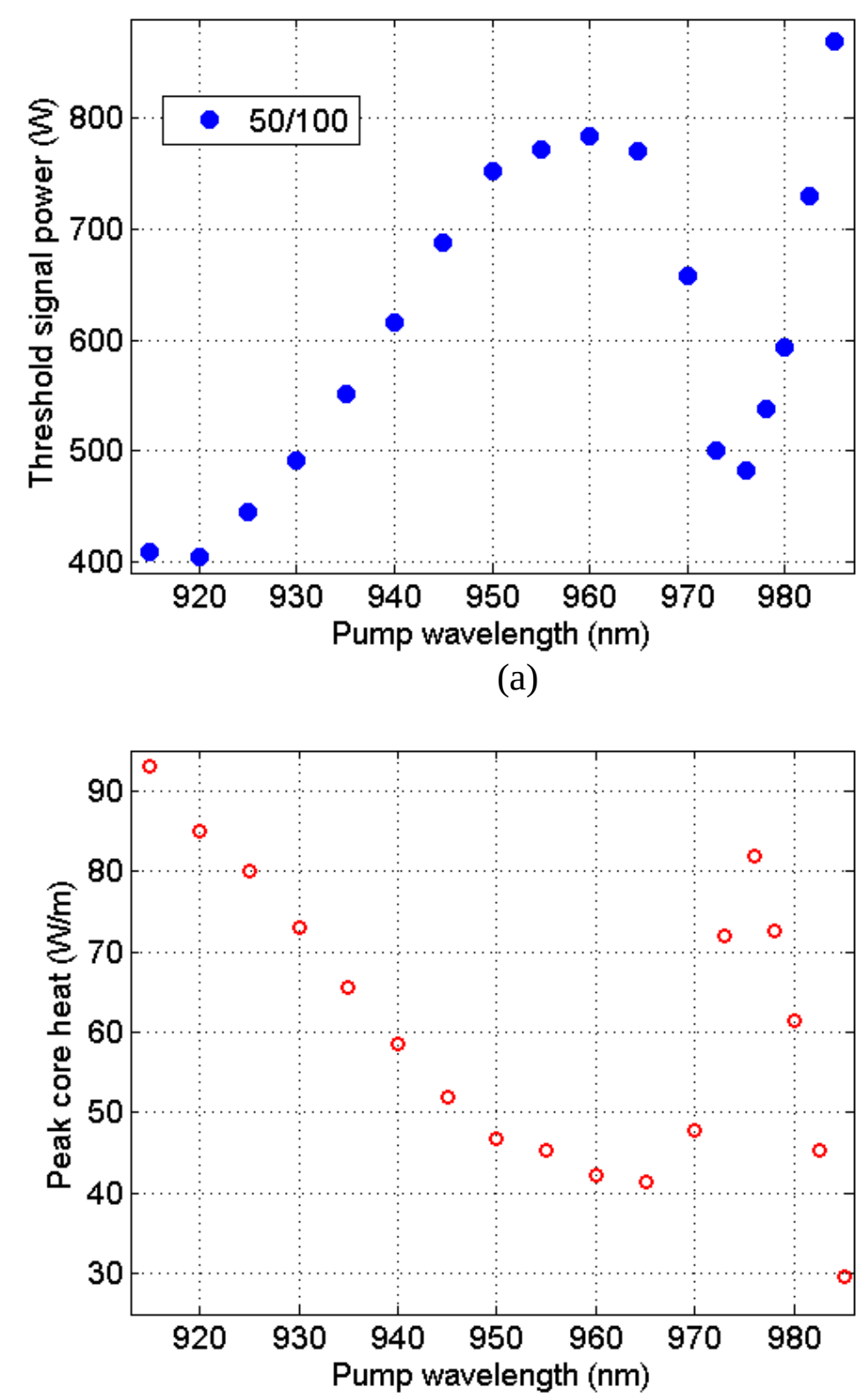

Fig. 2 Threshold power (a) and peak core heat (b) as a function of pump wavelength.

Intensity-noise-induced-MI thresholds as a function of signal wavelength are plotted in Fig. 3. The fiber parameters used in calculation are listed in Table 2, and those the same as in Table 1 are not listed. All the fiber is fully doped. Intensity noise of the signal was set to $R_N$=$10^{-10}$, which corresponds to a laser with high relative intensity noise and yields a realistic MI threshold [27]. It is shown that MI threshold is independent of rare earth dopant concentration [28]. To facilitate fast computation and save time, the length of the fiber was taken to be as short as 1m and the dopant concentrations of the ytterbium ions are adjusted the minimum value necessary to achieve high efficiency, defined as pump absorption >0.95, and to avoid overcompensate that of changing the pumping wavelength. However, if other length dependence nonlinear effects, such as stimulated Brillouin scattering, stimulated Raman scattering, should be taken into consideration, the length of the fiber must be chosen according to simulation condition.

TABLE II
PARAMETERS OF TEST AMPLIFIER

| | |
|---|---|
| $R_{core}$ | 10, 15μm |
| $R$ | 200μm |
| $n_{clad}$ | 1.45 |
| $NA$ | varies |
| $\xi_0$ | 0.01 |
| $R_N$ | $10^{-10}$ |
| *absorption* | 0 |
| $R_{coil}$ | ∞ |

As observed in [16], the results in Fig. 3 indicate clearly that the highest MI threshold can be obtained by operating at a seed wavelength of around 1032 nm, which is different from the theoretical results in [29]. In [29], MI threshold reduces monotonically with the increase of pump wavelength, which may due to that their method has not taken gain saturation or spatial hole burning into consideration. By shifting the seed signal to longer or shorter wavelengths, the MI threshold drops and the decrease in the longer wavelength is steeper than that in the shorter wavelength, which is also observed in [16]. The maximal MI threshold power at ~1030nm is due to the fact that the maximal signal emission coefficient is at ~1030nm, which lead to lower upper state populations, and tends to increase hole burning and thus increase threshold power. According to Eq. (2b), larger signal emission coefficient will lead to lower upper state populations and increasing of spatial hole burning, which results in the change in the transverse heat profile [25] and the increase of threshold power. The MI thresholds as a function of signal wavelength during 1055nm and 1075nm (as shown in the inset figure of Fig. 3) agree qualitatively with the experimental results [17], and some slight deviation may due to the difference of seed lasers.

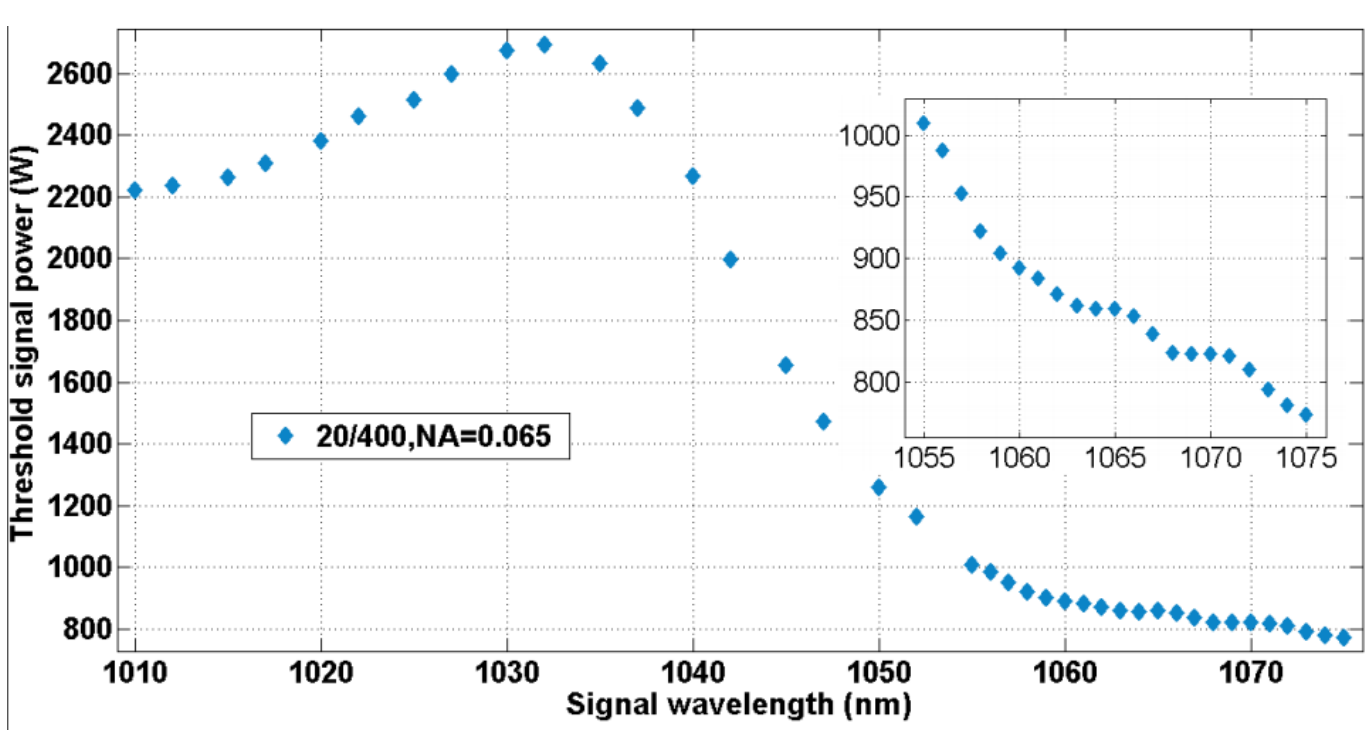

Fig. 3. Threshold power as a function of the signal wavelength for fiber with 20 μm diameter core and doping, 400 μm diameter pump cladding, and 0.065 core *NA*

For fiber with larger core size, such as 30μm as shown in Fig. 4(a), there is no maximal MI threshold power but a simple monotonic change. The monotonic change is expected due to the quantum defect changing as a function of signal wavelength, but the data shows structure that goes beyond this dependence. And the effect of the hole-burning is also obvious at near 1030nm. By reducing the *NA* as shown in Fig. 4(b), maximal MI threshold power at ~ 1030nm shows up again. It also reveals that MI threshold can be increased by reducing the *NA*. This may be due to that a lower value of the *NA* would allow the modes to expand slightly into the cladding [30], and $LP_{11}$ mode expands deeper than $LP_{01}$ mode [31], which results in that reduction of the overlap between $LP_{11}$ mode and dopant area is larger than that between $LP_{01}$ mode and dopant area. Similar to reducing the doping diameter to a smaller diameter than the core index step, reducing the *NA* can increase the threshold power. Figs. 3 and 4 also indicates that operating the amplifier in the shorter wavelengths can provide a substantial increase in the power threshold at which the onset of the modal instability phenomena occurs, as compared to operating the amplifier at longer wavelengths. For example, comparing the threshold results at 1030 nm to those at 1070 nm for 20/400 fiber, the threshold can be enhanced by a factor of 3.3.

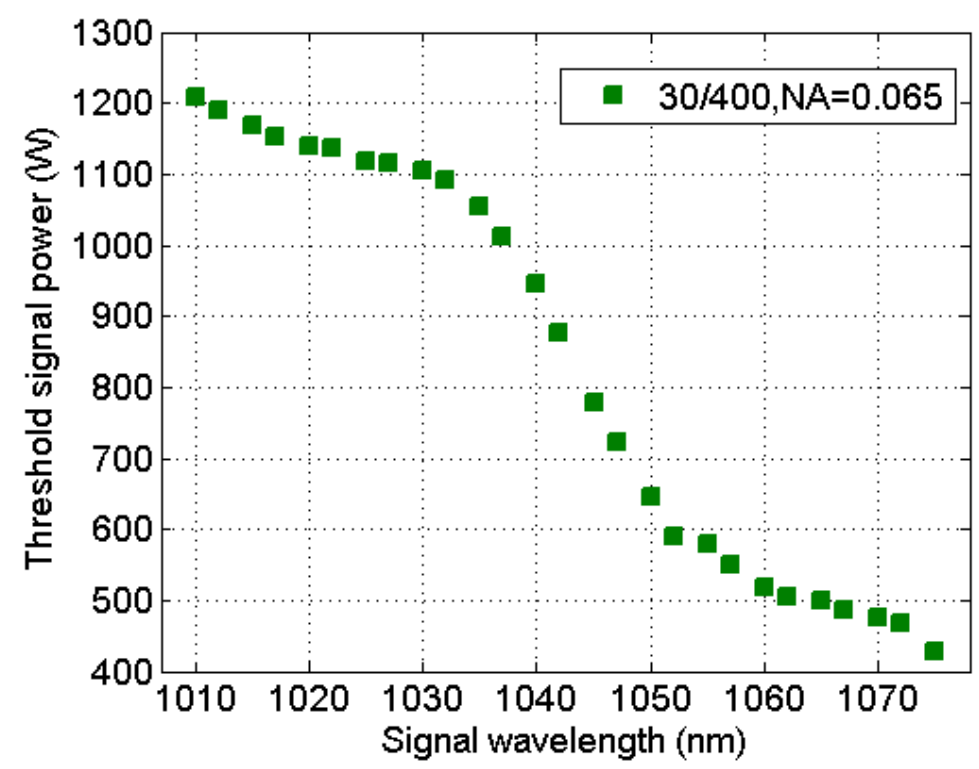

(a)

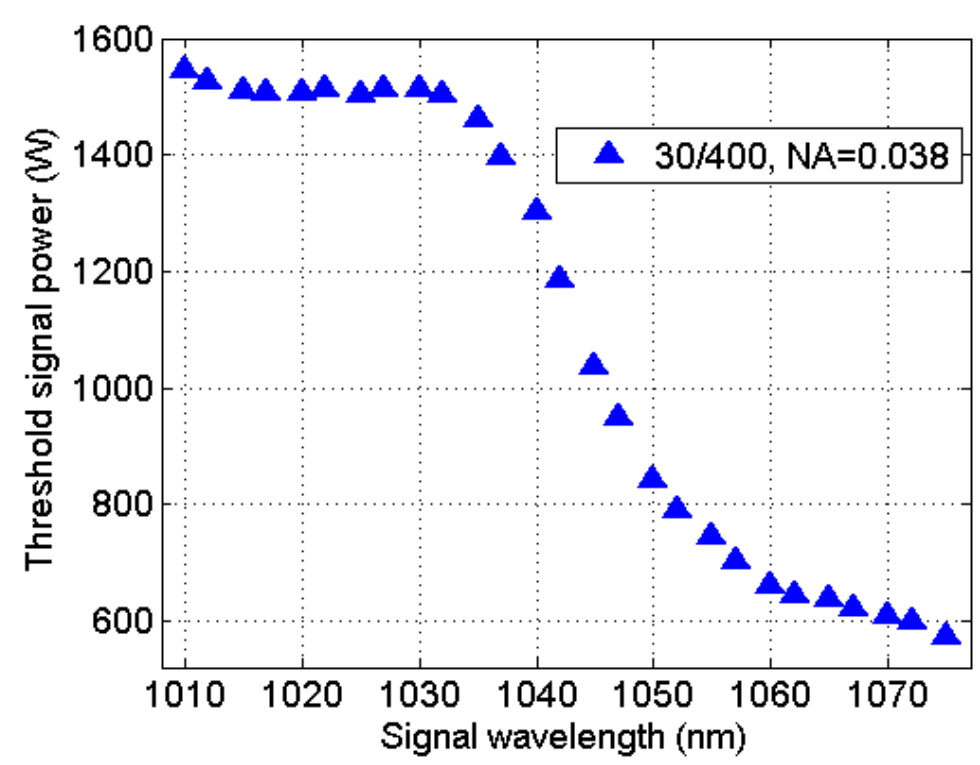

(b)

Fig. 4. Threshold power as a function of the signal wavelength for different type of fibers. (a) 30 μm diameter core and doping, 400 μm diameter pump cladding, 0.065 core *NA*, (b) 30 μm diameter core and doping, 400 μm diameter pump cladding, 0.038 core *NA*

It is theoretically predicted in [32, 33] that photodarkening can strongly reduce the threshold. The influence of photodarkening on MI as a function of wavelength is studied, which is shown in Fig. 5 (a). As pointed out in [34, 35], the photodarkening-induced loss is non-uniformly distributed along the length of the gain fiber as well as across the fiber core section. In the calculation, we employ a photodarkening model, which can account for the transverse shape and longitudinal variation of the absorption and is different from the model in [32, 33]. The longitudinal variation of photodarkening-induced loss is due to that the population inversion is different along the fiber length [35]. The photodarkening model was incorporated into our model through the linear absorption coefficient in Eq. (1). Due to that the high order mode content (<5%) is much smaller than the fundamental mode content (>95%) for the case in the model, the influence of high order mode content on transverse profile of the population inversion along the fiber is negligible. The transverse profile of the population inversion along the fiber was first computed in the presence of the fundamental mode alone through the irradiance based model as presented in [36]. And then the computed population inversion along the fiber was used to compute the photodarkening absorption strength. It is shown in [37] that the equilibrium photodarkening linearly depends on population inversion. So we calculated the equilibrium photodarkening losses by multiplying the average inversion level with 22.4dB/m, which is chosen to make the reduction in efficiency is ~1% at 1010nm. Finally, the calculated photodarkening absorption loss was used in Eq. (1) to investigate the influence of the photodarkening. It reveals that photodarkening reduces the MI threshold significantly at shorter wavelength, which is due to the higher photodarkening losses caused by the higher average remaining inversion [19], for example that the remaining inversion left at 1032nm is higher than that at 1070nm as shown in Fig. 5(b), and weakens the suppression of MI through shifting the seed wavelength. At wavelength longer than 1050nm, the influence of photodarkening on threshold power is relatively small, which is due to that the lower photodarkening losses induced by the lower remaining inversion left by laser operation at these wavelengths [19].

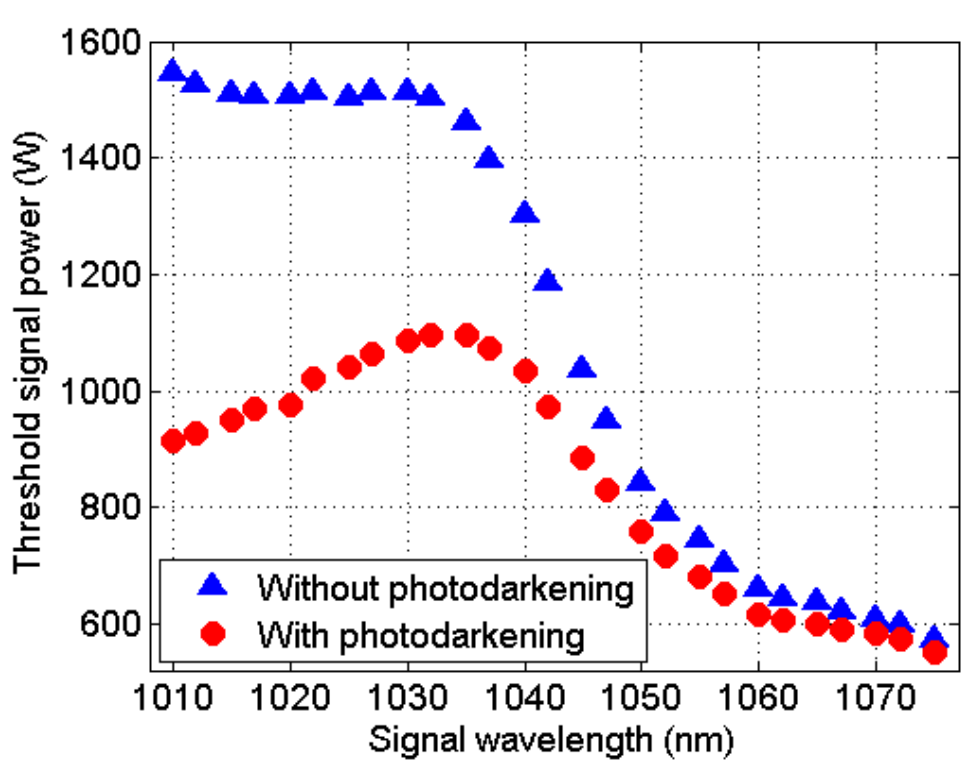

(a)

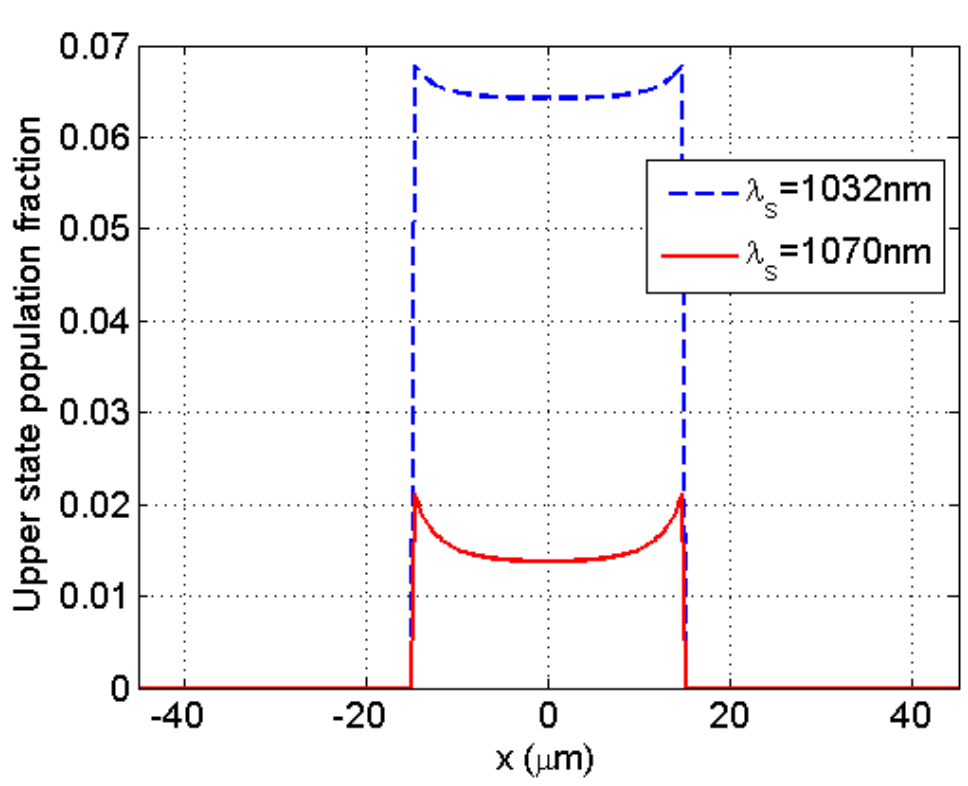

(b)

Fig. 5 The effect of the photodarkening on threshold power at different signal wavlengths (a), and the upper state population fraction at different wavelength at the middle of the fiber amplifier (b) for fiber with 30 μm diameter core and doping, 400 μm diameter pump cladding, 0.038 core *NA*.

## IV. Conclusions

In summary, based on an improved semi-analytical model, we presented theoretical study of wavelength dependence of mode instability. The model is compared with numerical model, and the model agrees well with that numerical model, with relative error less than 5%. It shows that, by shifting the pump wavelength from 976nm to 970nm or 985nm, the threshold power can be increased by 40% or 80%, respectively. MI threshold also shows a dependence on seeding wavelength, and the threshold can be increased by a factor of 3.3 through seeding at 1030 nm instead of 1070nm. The threshold was found to reduce strongly at shorter wavelength by photodarkening. Those results can be used to explain the experimental observation.